\documentclass[aps,pra,twocolumn,10pt,superscriptaddress,showpacs,amsmath,amssymb,nofootinbib]{revtex4-1}

\usepackage{graphicx}
\usepackage{hyperref}
\usepackage{array}
\usepackage{dcolumn}
\newcolumntype{L}[1]{>{\raggedright\let\newline\\\arraybackslash\hspace{0pt}}m{#1}}
\newcolumntype{C}[1]{>{\centering\let\newline\\\arraybackslash\hspace{0pt}}m{#1}}
\newcolumntype{R}[1]{>{\raggedleft\let\newline\\\arraybackslash\hspace{0pt}}m{#1}}

\newcommand{\postable}{
\begin{table*}[t]
\begin{ruledtabular}
\begin{tabular}{l L{13cm} c  r R{1.0cm} }
No.&POS question & Mean (stdev.) & $d_{\mathrm{T}}$ & $d_{\mathrm{R}}$ \\ \hline
1&I had a personal reason for choosing the research project I worked on.& 1.95 (1.02) & 1.25 & $1.06$ \\
2&My research project was interesting.& 1.40 (0.58) & 1.32 & $0.72$ \\
3&My research will help solve a problem in the world.& 2.20 (0.81) & 0.84 & $0.25$\\
4&I faced challenges that I managed to overcome in completing my research project.& 1.75 (0.70) & 0.77 & $0.24$ \\
5&My research project was exciting.& 1.95 (0.92) & 0.60 & $0.11$ \\
6&I was responsible for the outcome of my research.& 1.70 (0.46) & 0.54 & $0.08$ \\
7&The research question I worked on was important to me.& 2.30 (1.05) & 0.41 & $0.03$ \\
8&The findings of my research project gave me a sense of personal achievement.& 1.75 (0.62) & 0.82 & $-0.08$ \\
9&In conducting my research project, I actively sought advice and support.& 1.80 (0.81) & 0.51 & $-0.26$ \\
10&My findings were important to the scientific community.& 2.65 (0.73) & 0.24 & $-0.60$\\ \hline
11&To what extent does the word \emph{surprised} describe your experience in the labratory course?& 3.10 (0.94) & 0.16 & $0.05$ \\
12&To what extent does the word \emph{astonished} describe your experience in the labratory course?& 3.70 (0.95) & $-0.06$ & $-0.12$ \\
13&To what extent does the word \emph{amazed} describe your experience in the labratory course?& 3.00 (1.10) & $-0.05$ & $-0.34$ \\ 
14&To what extent does the word \emph{happy} describe your experience in the labratory course?& 2.60 (0.92) & 0.14 & $-0.65$ \\
15&To what extent does the word \emph{joyful} describe your experience in the labratory course?& 3.15 (0.85) & $-0.09$ & $-0.80$ \\
16&To what extent does the word \emph{delighted} describe your experience in the labratory course?& 3.15 (1.06) & $-0.20$ & $-0.82$ \\
\end{tabular}
\end{ruledtabular}
\caption{POS scores. Cohen $d$-values $d_\mathrm{T}$ and $d_\mathrm{R}$ compare POS scores from the Lasers Course to the Traditional and Research Courses, respectively. See text for interpretation of $d$-values.}
\label{tab:POS}
\end{table*}
}

\newcommand{\irrtable}{
\begin{table}
\begin{ruledtabular}
\begin{tabular}{l c c r}
Code & Agreement & Kappa & Interpretation \\ \hline
Personal Agency & 80\% & 0.61 & Substantial\\
Self-Efficacy & 75\% & 0.56 & Moderate\\
Peer Interactions & 84\% & 0.67 & Substantial\\
Affect & 91\% & 0.80 & Almost perfect
\end{tabular}
\end{ruledtabular}
\caption{Inter-rater reliability metrics for each coding categories across all 116 reflections.}
\label{tab:IRR}
\end{table}
}

\begin{document}

\title{Investigating student ownership of projects in an upper-division physics lab course}
\author{Jacob T. Stanley}
\affiliation{Department of Physics, University of Colorado, Boulder, CO 80309, USA}

\author{Dimitri R. Dounas-Frazer}
\affiliation{Department of Physics, University of Colorado, Boulder, CO 80309, USA}

\author{Laura Kiepura}
\affiliation{Department of Physics, Georgia State University, Atlanta, GA 30302, USA}

\author{H. J. Lewandowski}
\affiliation{Department of Physics, University of Colorado, Boulder, CO 80309, USA}
\affiliation{JILA, National Institute of Standards and Technology and University of Colorado, Boulder, CO, 80309, USA}


\begin{abstract}
In undergraduate research experiences, student development of an identity as a scientist is coupled to their sense of ownership of their research projects. As a a first step towards studying similar connections in physics laboratory courses, we investigate student ownership of projects in a lasers-based upper-division course. Students spent the final seven weeks of the semester working in groups on final projects of their choosing. Using data from the Project Ownership Survey and weekly student reflections, we investigate student ownership as it relates to students' personal agency, self-efficacy, peer interactions, and complex affective responses to challenges and successes. We present evidence of students' project ownership in an upper-division physics lab. Additionally, we find that there is a complex relationship between student affect and their sense of ownership.
\end{abstract}


\maketitle

\section{Introduction}
In the context of undergraduate research experiences, students' adoption of an identity as a scientist is connected, in part, to their level of ownership of their research project~\cite{Laursen2010}. However, not all students have the opportunity to participate in research experiences, and there is a need for other educational interventions in which students may develop their identities~\cite{Irving2015}. Laboratory courses are one potential space for these types of interventions---some recent work has addressed student ownership in biology and chemistry~\cite{Hanauer2012,Hanauer2014}. Here, we explore how student ownership manifests in an upper-division physics laboratory course.

Project ownership can be defined as a combination of a student's feeling of personal responsibility, level of commitment or buy-in, and sense of personal connection to the project~\cite{Hanauer2012,Wiley2009}. Furthermore, ownership is a byproduct of the interaction between a student and the environment~\cite{Hanauer2012}. The following characteristics have been identified as hallmarks of student ownership~\cite{Hanauer2014,Hanauer2012}: personal agency in the goal-setting process (\emph{intentionality}); confidence and willingness to contend with problems (\emph{self-efficacy)}; opportunity to work closely with other students (\emph{peer interaction}); expressions of excitement and a sense of personal achievement (\emph{positive affect}); and perception of the project as personally interesting and of value to the scientific community (\emph{relevance}). In the present work, we use this multifaceted interpretation of ownership.

Recently, tools have been developed to facilitate the study and design of upper-division physics laboratory courses by attending to student reasoning in, and attitudes about, experimental physics~\cite{Zwickl2015arXiv,Zwickl2014}. However, for existing instruments that explicitly address student ownership, we must look beyond the domain of physics education. One such instrument, the Project Ownership Survey (POS), was developed and validated for use in Course-based Undergraduate Research Experiences (CUREs) in biology~\cite{Hanauer2014}. We explored how the POS might be used in physics by implementing the survey in an upper-division physics laboratory course, which included a seven-week-long final project. These POS results were contextualize through a comparison to POS results obtained in both traditional and research-based biology laboratory courses~\cite{Hanauer2014}. Additional insight was provided by weekly reflections through which the physics students responded to prompts specifically designed to align with the above hallmarks of student ownership. We will show, the physics course fosters many hallmarks of project ownership among students, as indicated by a portion of the POS and reflection data. However, while the positive emotive scales of the POS have been useful measures of project ownership in other contexts~\cite{Hanauer2014}, students in our study described complex affect to the struggles and successes that accompanied their final projects, making it difficult to interpret responses to emotive survey items.

\vspace{-10pt}

\section{Laboratory Course Context}
We studied an upper-division physics laboratory course, the ``Lasers Course," which focused on topics in contemporary optics and enrolled 20 students. This course was offered at private, primarily undergraduate institution with a total enrollment of 6,500 students, 7.7\% of whom are underrepresented minorities and 62\% of whom are women. As articulated by the instructor, course learning goals included development of students' technical skills, conceptual understanding, and appreciation of the importance of lasers and optics in science and industry. Other major goals included fostering students' feelings of confidence in their experimental physics abilities, excitement about experimental physics, and ownership over their final projects. The course lasted 14 weeks and was split into two halves. During the first half, students engaged in guided experimental skill-building activities. During the second half, students worked on final projects in groups of 2--4 people. Students were presented with a portfolio of possible projects and ranked each project according to their interest. Groups were formed based on student preferences for project, and each student was assigned to their first or second choice.

In total, there were six distinct projects, all of which were informed by the instructor's own research. As an example, one project was the continued development of a frequency comb---a useful tool for measuring optical frequencies found in many modern atomic, molecular, and optical physics experiments. In contrast to the first half of the course, there were no lab guides that students could reference for guidance on their projects. Instead, students were provided with relevant scientific journal articles and had frequent access to their instructor as an additional resource. There was no official time during which students were required to be in the lab; groups coordinated their own plans for time management and division of labor. Each student spent roughly 10--15 hours per week working on their projects. Projects culminated in oral presentations and a report written in the style of a typical journal article. Throughout the course, students completed weekly reflections on their progress.

\vspace{-10pt}

\section{Methods}
We collected interview data, observational data, survey data, and course artifacts from students enrolled in the Lasers Course. All 20 students (3 women, 17 men) who completed the course participated in the study. Herein, we limit our discussion to POS data and students' reflections on their final projects.

To contextualize our POS results, we compared our data to those reported by Hanauer~\emph{et al.} (2014) for students enrolled in traditional and research-based laboratory courses (hereafter, the ``Traditional" and ``Research Courses"). All of the comparison courses were biology courses taught by instructors who were members of the CURE Network~\cite{Hanauer2014}. While comparison across disciplines is not ideal, we are not aware of existing data in the literature on ownership in upper-division physics laboratory courses, which might facilitate intra-disciplinary comparisons. Traditional Courses were ``described by students as being a required introductory course and involving various small experiments; not real research," whereas Research Courses were ``described by students in terms of the scientific question they explored" (Hanauer, 2014; p.156). 

We administered the POS during the final week of the Lasers Course. All 20 students completed the survey. The POS consists of 16 five-point Likert items, where responses range from Strongly Agree~(1) to Strongly Disagree~(5). For each survey item, we computed the mean and standard deviation of student responses. To compare our POS results to those obtained in the comparison courses, we used the means and standard deviations reported in Ref.~\cite{Hanauer2014} to compute Cohen's $d$-value for effect sizes~\cite{Cohen1992}. For each survey item, we computed two $d$-values, $d_{\mathrm{T}}$ and $d_{\mathrm{R}}$, which compare the Lasers Course to the Traditional and Research Courses, respectively.

\irrtable

To gain additional insight, we analyzed weekly reflections completed by students during the project portion of the Lasers Course. The research team collaborated with the instructor to develop reflection prompts that would elicit information about students' \emph{intentionality}, \emph{self-efficacy}, \emph{peer interactions}, and \emph{affect} during their projects. Three sets of prompts were created: one focused on weekly goals; another on technical problems; and the third focused on successful moments. Students were prompted to describe how they felt and to summarize their own role, the role of their peers, and the role of their instructor in navigating these experiences. None of the prompts were designed to probe \emph{relevance} in order to avoid the possibility of repeatedly calling attention to lack of relevance for students who may not have found their project interesting or valuable. The modified reflections were administered during the last six weeks of the seven-week project. Prompts were alternated from week to week such that each set was administered twice.

\postable

Reflections were coded for experiences that demonstrated \emph{intentionality}, \emph{self-efficacy}, \emph{peer interactions}, and/or \emph{affect} according to the following operational definitions:
\begin{quote}
\emph{Intentionality}: The student described playing an active role in the goal-setting process.\\[6pt]
\emph{Self-efficacy}: The student described a willingness and confidence to contend with problems they encountered during the project.\\[6pt]
\emph{Peer interaction}: The student described an episode where a peer provided technical, strategic, conceptual, or emotional support.\\[6pt]
\emph{Affect}: The student used any emotive language.
\end{quote}
Because many students alternated between singular and plural personal pronouns (``I" and ``we") in their reflections, we did not distinguish between personal or collective intentionality or efficacy. The operational definition of \emph{affect} is broad by design: we wanted to avoid categorizing emotive language according to any \emph{a priori} scheme so that we may instead try to understand students' affective responses in connection to their \emph{intentionality}, \emph{self-efficacy}, and \emph{peer interactions}~\cite{Schutz2002}.

In addition to these definitions, our coding scheme consisted of examples and counterexamples from our dataset. The coding scheme was developed over three iterations of independent coding by two of the authors (J.T.S. and D.R.D.F.). The final scheme was applied to the entire dataset independently by the same two raters. Discrepancies were resolved through discussion with the research team as a whole. To determine the inter-rater reliability of our coding, scheme we calculated percentage agreement and Cohen's kappa statistic for all 116 reflections. The results and their interpretations are shown in Table \ref{tab:IRR}.

\vspace{-10pt}

\section{Results and Discussion}

POS results and Cohen's $d$-values are shown in Table~\ref{tab:POS}. We divided POS results into two sections, informed by the survey design~\cite{Hanauer2014}: (1) items 1--10, which were designed to address degrees of project ownership; and (2) items 11--16, which represent several emotive scales. Within each section, survey items are listed in order of decreasing $d_\mathrm{R}$. For a given survey item, the sign of the $d$-value corresponds to whether there is more (positive) or less (negative) agreement with the survey item among students in the Lasers Course than in the comparison courses. The magnitude is proportional to the size of the disparity: small, medium, and large differences correspond to $d$-values of approximately 0.2, 0.5, and 0.8, respectively~\cite{Cohen1992}. We interpret positive $d_\mathrm{R}$ values as indicative of high levels of ownership and negative $d_\mathrm{T}$ values as indicative of low levels of ownership.

POS responses on items 1--10 are generally indicative of high levels of ownership in the Lasers Course: the mean Likert responses vary from 1.40--2.65, indicating general agreement with the items;  $d_\mathrm{T}$ varies from 0.24--1.32, corresponding to small-to-large positive differences between the Lasers Course and the Traditional Course; and, in 7 of the 10 items, $d_\mathrm{R}$ varies between 0.03--1.06, corresponding to small-to-large positive differences compared to the Research Courses. However, the situation is markedly different for items 11--16: the mean Likert responses vary from 2.60--3.70, indicating overall neutral responses to the items; the magnitude of $d_\mathrm{T}$ is 0.2 or smaller, corresponding to only small differences between the Lasers Course and the Traditional Courses; and, in 5 of the 6 items, $d_\mathrm{R}$ represents small-to-large negative differences compared to the Research Courses. Thus, the POS responses communicate two conflicting messages: taken as a group, items 1--10 demonstrate high levels of ownership, whereas the emotive scales (items 11--16) do not. While these results seem to be in tension, they are nevertheless consistent with the experiences articulated in student reflections. 

The student reflections highlight multiple examples of \emph{intentionality}, \emph{self-efficacy}, and \emph{peer interactions}, as demonstrated by the following excerpts from student reflections. For example:
\begin{itemize}
\item[]
\emph{``I feel that I have helped efficiently plan out and articulate goals for progress, I have at some times been able to `pop off ideas' to help get us over a `hump' when we're stuck or frustrated \ldots"}
\end{itemize}
This quote is an example of \emph{intentionality} because the student describes that they were partly responsible for goal-setting and brainstorming ideas. We found that 70\% of students described at least one instance of \emph{intentionality} over the course of the project. Given that personal responsibility is coupled to one's personal agency in goal-setting \cite{Wiley2009}, the evidence of \emph{intentionality} in the reflections complements the level of agreement with POS item 6 (``I was responsible for the outcome of my research").

In the following example, a student describes changes in their \emph{self-efficacy}:
\begin{itemize}
\item[]
\emph{``Now that we have made some kind of progress we're less stressed because we know we have at least something to show for all of our work, and we are accomplishing things faster. After being stuck for so long feelings of not ever being able to solve the problem were creeping in. Now it feels like we can get through any problem we encounter."}
\end{itemize}
Here, the student described how a long period of feeling ``stuck" on their project caused them to begin doubting their ability to contend with problems. However, after making progress on the project, they felt confident that they could handle ``any problem we encounter." This example, which is typical of the efficacious statements in our dataset, aligns with established connections between mastery experiences and development of self-efficacy beliefs~\cite{Bandura1996}. Overall, 80\% of students described positive \emph{self-efficacy} at least twice during the project, a finding which is coupled to the level of agreement with POS item~4 (``I faced challenges that I managed to overcome \ldots").

Students described several types of \emph{peer interaction}. Strategic planning was one common theme:
\begin{itemize}
\item[]
\emph{``We had a meeting were we talked about everything that needed to get done. We prioritized the things we could possibly do \ldots"}
\end{itemize}
In this example, the student strategized with their group about next steps on the project. We found 85\% of students described at least three instances of \emph{peer interaction}, which reinforces the the level of agreement with POS item 9 (``\ldots I actively sought advice and support").

Expressions of \emph{intentionality}, \emph{self-efficacy}, and \emph{peer interaction} in the reflections are commensurate with the responses to items 1--10 of the POS. On the other hand, while Likert responses to the emotive scales (items 11--16) are generally neutral, we nevertheless see a range of affective responses in students' reflections. For example, one student described complex emotional reactions to a successful moment on their project:
\begin{itemize}
\item[]
\emph{``Not having a mode lock delayed progress a couple weeks. Putting time into something that was not giving many signs of life, that wasn't giving clear directions of where to look next, was frustrating and draining. So that makes this case of finding one an ecstatic occasion \ldots''}
\end{itemize}
In this excerpt, the student described feeling uncertain during a period of time when they weren't seeing much progress on their project, a common experience among scientists~\cite{Hanauer2012,Jaber2014}. The student further described feeling frustrated and drained during this time. However, upon seeing a clear sign of progress (\emph{e.g.}, finding a mode lock), the student expressed excitement. This suggests that the emotional evolution of students' experience in the lab setting is complex and does not fit neatly into one particular affective category. 

The nature of students' articulated affective experiences makes it difficult to know how students interpreted emotive scales on the POS. Some students may have responded according to whether or not they ever experienced a particular emotion during the course. Others may have ``integrated" their emotional experiences over time, responding based on the overall balance of ``negative'' and ``positive'' experiences. Still further, some students' responses may have been based on their most recent emotional experience. Thus, in the context of upper-division physics laboratory courses, there is a need for continued investigation of the nature of students' affective experiences and the connection to project ownership.

\vspace{-10pt}

\section{Summary and Future Directions}
We have shown the students in this Lasers Course have a high level of agreement with many of the hallmarks of ownership that have been outlined in previous works. We found similar connections between ownership, intentionality, self-efficacy, and peer interaction in the physics context that were similar to those found in a Biology lab context. On the other hand, the complex affective dynamics we saw, warrant further investigation. Future work will focus on exploring the role of these and other dynamics---such as group interactions---in fostering ownership in upper-division physics laboratory courses.

\vspace{-10pt}

\section{Acknowledgements}
We acknowledge B. Wilcox and A. Little for useful discussions. This work was supported by NSF grant nos. DUE-1323101 and DUE 133-4170.

\bibliography{./Ownership}

\begin{thebibliography}{11}%
\makeatletter
\providecommand \@ifxundefined [1]{%
 \@ifx{#1\undefined}
}%
\providecommand \@ifnum [1]{%
 \ifnum #1\expandafter \@firstoftwo
 \else \expandafter \@secondoftwo
 \fi
}%
\providecommand \@ifx [1]{%
 \ifx #1\expandafter \@firstoftwo
 \else \expandafter \@secondoftwo
 \fi
}%
\providecommand \natexlab [1]{#1}%
\providecommand \enquote  [1]{``#1''}%
\providecommand \bibnamefont  [1]{#1}%
\providecommand \bibfnamefont [1]{#1}%
\providecommand \citenamefont [1]{#1}%
\providecommand \href@noop [0]{\@secondoftwo}%
\providecommand \href [0]{\begingroup \@sanitize@url \@href}%
\providecommand \@href[1]{\@@startlink{#1}\@@href}%
\providecommand \@@href[1]{\endgroup#1\@@endlink}%
\providecommand \@sanitize@url [0]{\catcode `\\12\catcode `\$12\catcode
  `\&12\catcode `\#12\catcode `\^12\catcode `\_12\catcode `\%12\relax}%
\providecommand \@@startlink[1]{}%
\providecommand \@@endlink[0]{}%
\providecommand \url  [0]{\begingroup\@sanitize@url \@url }%
\providecommand \@url [1]{\endgroup\@href {#1}{\urlprefix }}%
\providecommand \urlprefix  [0]{URL }%
\providecommand \Eprint [0]{\href }%
\providecommand \doibase [0]{http://dx.doi.org/}%
\providecommand \selectlanguage [0]{\@gobble}%
\providecommand \bibinfo  [0]{\@secondoftwo}%
\providecommand \bibfield  [0]{\@secondoftwo}%
\providecommand \translation [1]{[#1]}%
\providecommand \BibitemOpen [0]{}%
\providecommand \bibitemStop [0]{}%
\providecommand \bibitemNoStop [0]{.\EOS\space}%
\providecommand \EOS [0]{\spacefactor3000\relax}%
\providecommand \BibitemShut  [1]{\csname bibitem#1\endcsname}%
\let\auto@bib@innerbib\@empty
\bibitem [{\citenamefont {Laursen}\ \emph {et~al.}(2010)\citenamefont
  {Laursen}, \citenamefont {Hunter}, \citenamefont {Seymour}, \citenamefont
  {Thiry},\ and\ \citenamefont {Melton}}]{Laursen2010}%
  \BibitemOpen
  \bibfield  {author} {\bibinfo {author} {\bibfnamefont {S.}~\bibnamefont
  {Laursen}}, \bibinfo {author} {\bibfnamefont {A.-B.}\ \bibnamefont {Hunter}},
  \bibinfo {author} {\bibfnamefont {E.}~\bibnamefont {Seymour}}, \bibinfo
  {author} {\bibfnamefont {H.}~\bibnamefont {Thiry}}, \ and\ \bibinfo {author}
  {\bibfnamefont {G.}~\bibnamefont {Melton}},\ }\href@noop {} {\emph {\bibinfo
  {title} {Undergraduate {R}esearch in the {S}ciences}}}\ (\bibinfo
  {publisher} {Jossey-Bass},\ \bibinfo {year} {2010})\BibitemShut {NoStop}%
\bibitem [{\citenamefont {Irving}\ and\ \citenamefont
  {Sayre}(ress)}]{Irving2015}%
  \BibitemOpen
  \bibfield  {author} {\bibinfo {author} {\bibfnamefont {P.~W.}\ \bibnamefont
  {Irving}}\ and\ \bibinfo {author} {\bibfnamefont {E.~C.}\ \bibnamefont
  {Sayre}},\ }\href@noop {} {\bibfield  {journal} {\bibinfo  {journal} {Journal
  of the Scholarship of Teaching and Learning}\ } (\bibinfo {year} {{in
  press}})}\BibitemShut {NoStop}%
\bibitem [{\citenamefont {Hanauer}\ \emph {et~al.}(2012)\citenamefont
  {Hanauer}, \citenamefont {Frederick}, \citenamefont {Fotinakes},\ and\
  \citenamefont {Strobel}}]{Hanauer2012}%
  \BibitemOpen
  \bibfield  {author} {\bibinfo {author} {\bibfnamefont {D.~I.}\ \bibnamefont
  {Hanauer}}, \bibinfo {author} {\bibfnamefont {J.}~\bibnamefont {Frederick}},
  \bibinfo {author} {\bibfnamefont {B.}~\bibnamefont {Fotinakes}}, \ and\
  \bibinfo {author} {\bibfnamefont {S.~A.}\ \bibnamefont {Strobel}},\ }\href
  {\doibase 10.1187/cbe.12-04-0043} {\bibfield  {journal} {\bibinfo  {journal}
  {CBE Life Sciences Education}\ }\textbf {\bibinfo {volume} {11}},\ \bibinfo
  {pages} {378} (\bibinfo {year} {2012})}\BibitemShut {NoStop}%
\bibitem [{\citenamefont {Hanauer}\ and\ \citenamefont
  {Dolan}(2014)}]{Hanauer2014}%
  \BibitemOpen
  \bibfield  {author} {\bibinfo {author} {\bibfnamefont {D.~I.}\ \bibnamefont
  {Hanauer}}\ and\ \bibinfo {author} {\bibfnamefont {E.~L.}\ \bibnamefont
  {Dolan}},\ }\href {\doibase 10.1187/cbe.13-06-0123} {\bibfield  {journal}
  {\bibinfo  {journal} {CBE Life Sciences Education}\ }\textbf {\bibinfo
  {volume} {13}},\ \bibinfo {pages} {149} (\bibinfo {year} {2014})}\BibitemShut
  {NoStop}%
\bibitem [{\citenamefont {Wiley}(2009)}]{Wiley2009}%
  \BibitemOpen
  \bibfield  {author} {\bibinfo {author} {\bibfnamefont {J.}~\bibnamefont
  {Wiley}},\ }\emph {\bibinfo {title} {{Student Ownership of Learning: An
  Analysis}}},\ \href@noop {} {\bibinfo {type} {Master's thesis}},\ \bibinfo
  {school} {University of Hawai'i} (\bibinfo {year} {2009})\BibitemShut
  {NoStop}%
\bibitem [{\citenamefont {Zwickl}\ \emph {et~al.}()\citenamefont {Zwickl},
  \citenamefont {Hu}, \citenamefont {Finkelstein},\ and\ \citenamefont
  {Lewandowski}}]{Zwickl2015arXiv}%
  \BibitemOpen
  \bibfield  {author} {\bibinfo {author} {\bibfnamefont {B.~M.}\ \bibnamefont
  {Zwickl}}, \bibinfo {author} {\bibfnamefont {D.}~\bibnamefont {Hu}}, \bibinfo
  {author} {\bibfnamefont {N.}~\bibnamefont {Finkelstein}}, \ and\ \bibinfo
  {author} {\bibfnamefont {H.~J.}\ \bibnamefont {Lewandowski}},\ }\href@noop {}
  {}\bibinfo {note} {{arXiv:1410.0881}}\BibitemShut {NoStop}%
\bibitem [{\citenamefont {Zwickl}\ \emph {et~al.}(2014)\citenamefont {Zwickl},
  \citenamefont {Hirokawa}, \citenamefont {Finkelstein},\ and\ \citenamefont
  {Lewandowski}}]{Zwickl2014}%
  \BibitemOpen
  \bibfield  {author} {\bibinfo {author} {\bibfnamefont {B.~M.}\ \bibnamefont
  {Zwickl}}, \bibinfo {author} {\bibfnamefont {T.}~\bibnamefont {Hirokawa}},
  \bibinfo {author} {\bibfnamefont {N.}~\bibnamefont {Finkelstein}}, \ and\
  \bibinfo {author} {\bibfnamefont {H.~J.}\ \bibnamefont {Lewandowski}},\
  }\href {\doibase 10.1103/PhysRevSTPER.10.010120} {\bibfield  {journal}
  {\bibinfo  {journal} {Phys. Rev. ST Phys. Educ. Res.}\ }\textbf {\bibinfo
  {volume} {10}},\ \bibinfo {pages} {010120} (\bibinfo {year}
  {2014})}\BibitemShut {NoStop}%
\bibitem [{\citenamefont {Cohen}(1992)}]{Cohen1992}%
  \BibitemOpen
  \bibfield  {author} {\bibinfo {author} {\bibfnamefont {J.}~\bibnamefont
  {Cohen}},\ }\href {\doibase 10.1037/0033-2909.112.1.155} {\bibfield
  {journal} {\bibinfo  {journal} {Psychological Bulletin}\ }\textbf {\bibinfo
  {volume} {112}},\ \bibinfo {pages} {155} (\bibinfo {year}
  {1992})}\BibitemShut {NoStop}%
\bibitem [{\citenamefont {Schutz}\ and\ \citenamefont
  {DeCuir}(2002)}]{Schutz2002}%
  \BibitemOpen
  \bibfield  {author} {\bibinfo {author} {\bibfnamefont {P.~A.}\ \bibnamefont
  {Schutz}}\ and\ \bibinfo {author} {\bibfnamefont {J.~T.}\ \bibnamefont
  {DeCuir}},\ }\href {\doibase 10.1207/S15326985EP3702\_7} {\bibfield
  {journal} {\bibinfo  {journal} {Educational Psychologist}\ }\textbf {\bibinfo
  {volume} {37}},\ \bibinfo {pages} {125} (\bibinfo {year} {2002})}\BibitemShut
  {NoStop}%
\bibitem [{\citenamefont {Bandura}\ \emph {et~al.}(1996)\citenamefont
  {Bandura}, \citenamefont {Barbaranelli}, \citenamefont {Caprara},\ and\
  \citenamefont {Pastorelli}}]{Bandura1996}%
  \BibitemOpen
  \bibfield  {author} {\bibinfo {author} {\bibfnamefont {A.}~\bibnamefont
  {Bandura}}, \bibinfo {author} {\bibfnamefont {C.}~\bibnamefont
  {Barbaranelli}}, \bibinfo {author} {\bibfnamefont {G.~V.}\ \bibnamefont
  {Caprara}}, \ and\ \bibinfo {author} {\bibfnamefont {C.}~\bibnamefont
  {Pastorelli}},\ }\href {\doibase 10.2307/1131888} {\bibfield  {journal}
  {\bibinfo  {journal} {Child Development}\ }\textbf {\bibinfo {volume} {67}},\
  \bibinfo {pages} {1206} (\bibinfo {year} {1996})}\BibitemShut {NoStop}%
\bibitem [{\citenamefont {Jaber}(2014)}]{Jaber2014}%
  \BibitemOpen
  \bibfield  {author} {\bibinfo {author} {\bibfnamefont {L.~Z.}\ \bibnamefont
  {Jaber}},\ }\emph {\bibinfo {title} {{Affective dynamics of students'
  disciplinary engagement in science}}},\ \href@noop {} {\bibinfo {type}
  {{Ph.D.} thesis}},\ \bibinfo  {school} {Tufts University} (\bibinfo {year}
  {2014})\BibitemShut {NoStop}%
\end{thebibliography}%

\end{document}